\documentclass{article}
\usepackage{attrib}
\usepackage{amsmath, amssymb, amsthm}
\newtheorem{prop}{Proposition}
\newtheorem{defin}{Definition}
\newtheorem{assum}{Assumption}
\usepackage[letterpaper, portrait, margin=1.25in]{geometry}
\usepackage{amsmath, amssymb}
\title{Culture, Computation, Morality}
\author{Jongmin Jerome Baek\\UC Berkeley}
\begin{document}
\maketitle

\abstract{
I point to a deep and unjustly ignored relation between culture and computation. First, I interpret Piaget's theory of child development with the language of theoretical computer science. Then I argue that the two different possible manifestations Piagetian disequilibrium are equivalent to two distinct cultural tendencies (what is commonly referred to as the East-West divide). I argue that this simple characterization of overaccommodation versus overassimilation provides a satisfying explanation as to why the two cultural tendencies differ in the way they empirically do. All such notions are grounded on a firm mathematical framework for those who prefer the computable, and grounded on my personal history for those who prefer the uncomputable.
}

\newpage
\section{Preppendix}
\emph{Proof of proposition: a human $H$ is not free if $H$ computes a free human $H'$.}
\begin{assum} The Church-Turing Thesis is true: everything that is physically computable is computable by some Turing machine. \end{assum}
\begin{defin} A human $H$ is a thing that does computation, i.e. an automaton, and is in the physical world. \end{defin}
\begin{defin} A human $H$ is free if and only if $H$ is uncomputable.\end{defin}
\emph{Corollary:} A human $H$ is not free if and only if $H$ is computable.
\begin{prop} A human $H$ is at most Turing-complete. \end{prop}
\emph{Proof:} This follows from the Church-Turing Thesis. 
\begin{prop} There exists no Turing machine $M$ that computes the output of an arbitrary Turing machine $J$.\end{prop}
\emph{Proof:} This follows from the undecidability of the halting problem.
\begin{prop} A human $H$ cannot compute an uncomputable function. \end{prop}
\emph{Proof:} By Proposition 1, $H$ cannot compute any function no Turing machine can compute. No Turing machine can compute an uncomputable function. Therefore $H$ cannot compute an uncomputable function.
\begin{defin} An automaton $M$ is ``stronger" than an automaton $G$ if the functions $G$ can compute is a strict subset of the functions $M$ can compute. Similarly, $M$ is ``stronger" than $G$ if and only if $G$ is ``weaker" than $M$. \end{defin}
\begin{prop} For some automaton $M$, if $M$ computes the output of an arbitrary Turing machine $J$, $M$ is either stronger than or weaker than a Turing-complete machine.\end{prop}
\emph{Proof:} By Proposition 2, no Turing machine $M$ computes the output of an arbitrary Turing machine $J$. Therefore, if $M$ computes the output of an arbitrary Turing machine, $M$ is not a Turing-complete machine. Therefore, $M$ is either stronger or weaker than a Turing-complete machine.
\begin{prop} If a human $H$ computes the output of an arbitrary Turing machine $J$, $H$ is weaker than a Turing-complete machine.\end{prop}
\emph{Proof:} By Proposition 4, if an automaton $H$ computes the output of an arbitrary Turing machine $J$, $H$ is either stronger or weaker than a Turing-complete machine. By Proposition 1, a human $H$ is no stronger than a Turing-complete machine. Therefore, $H$ is weaker than a Turing-complete machine. 
\begin{prop} If a human $H$ computes the output of an arbitrary Turing machine $J$, $H$ is computable by some Turing machine. \end{prop}
\emph{Proof:} By Proposition 5, if a human $H$ computes the output of an aribtrary Turing machine $J$, $H$ is less powerful than a Turing-complete machine.

\emph{Lemma 1:} There exists a Turing machine that can compute the outcome of any sub-Turing-complete machine.

By Lemma 1, if $H$ is less powerful than a Turing-complete machine, $H$ is computable by some Turing machine.

\begin{prop} If a human $H$ computes the output of an arbitrary Turing machine $J$, $H$ is not free. \end{prop}
\emph{Proof:} By Proposition 6, if a human $H$ computes the output of an arbitrary Turing machine $J$, $H$ is computable by some Turing machine. By the corollary to Definition 2, if $H$ is computable, $H$ is not free.
\begin{prop} If a human $H$ computes a free human $H'$, $H$ is not free. \end{prop}
\emph{Proof:} By Definition 2, a human $H'$ is free if and only if $H'$ is uncomputable. By Proposition 2, $H'$ is at most Turing-complete. Because $H'$ is uncomputable, $H'$ must be Turing-complete. To compute the output of a Turing-complete machine is tantamount to computing the output of an arbitrary Turing machine. By Proposition 7, if a human $H$ computes the output of an aribtrary Turing machine $J$, $H$ is not free. Therefore, if a human $H$ computes the output of a Turing-complete machine $H'$, $H$ is not free. Therefore, if a human $H$ computes a free human $H'$, $H$ is not free.
\newpage
\section{Preface}
When I attended kindergarten in the USA, I drew comic books about a stickman and a penguin fighting evil. Every time I finished drawing one I showed it to my teacher, who always took her time to flip through it and said things like ``Wow!", ``Very good!". Now, my parents are very loving people, probably the best parents anyone could ask for. However, or therefore, when they found out about this arrangement, they said, in Korean, ``How your teacher must be so sick and tired of having to pretend to enjoy your stupid comic books!" I had no reason to doubt my teacher's encouragements were genuine, and at the same time I had no reason to doubt my parents' comments about my teacher. Not that I had a theory to explain how the contradiction could be resolved; I just knew instinctively that both were true. Resolving this paradox took me fifteen years and believing that I and them and everyone else is a computer.

Let me explain. I was born in Seoul, South Korea twenty years ago. I lived four years there, then went to the alien land of America. I puked a lot for two years there, trying to figure out how these aliens -- Americans -- think. I then went to the newly alien land of Korea and sulked for six years, trying to figure out how these aliens -- Koreans -- think. I then again went to the again alien land of America, and for the past eight years, I have been running around back and forth between the two always alien homes, trying to figure out how both peoples think. To live with one foot in one culture and another foot in another is to constantly feel like your crotch will split in half.\footnote{``To feel like your crotch will split in half" is a clich{\`e}d Korean idiom.} If to live in one culture is to look at a giant intricate wooden totem with clear vision from the front, to live in two is to look at the same totem with horribly blurry vision from the front and the back. I squinted hard trying to figure out how the pattern on the front connects with the utterly contradictory pattern on the back. All existing theories I found were either feel-good and flaccid, or complete and condescending.\footnote{As a side effect of this search, I have developed a horrible revulsion towards academic psychology's tendencies to examine how person from a typically Western cultures think and declare it ``universal human nature". For an incisive critique, see \cite{Henrich2010a}.} I craved a method of synthesis, a single coherent theory that would explain the two contradictory patterns at once without putting one above the other. 

Then one day I stumbled upon this strange hum emanating from three siblings: Cantor's orders of infinity, G{\"o}del's Incompleteness theorem, and Turing's halting problem. The three siblings, while different in appearance, say together one exact same very intuitive thing: there is a limit\footnote{Or, as some may argue, an unexpected power} to logical reasoning. They also say a bit less intuitive thing: we can use logic to discover precisely what kinds of things there are that we cannot use logic to discover. In other words, \emph{we can know exactly what it is that we cannot know}. In another words, we can \emph{prove that we cannot prove} some things. The essence of this argument lies in the fact that self-reference, like the sentence ``this sentence is false", inevitably leads to a contradiction.\footnote{Douglas Hofstadter's poetry-laden cult classic tome \emph{G{\"o}del, Escher, Bach}\cite{geb} is probably the most famous book on this topic. As he puts it, the Incompleteness theorem says that ``for every record player, there are records which it cannot play because it will cause its indirect self-destruction." Hofstadter claims that G{\"o}del's Incompleteness theorem, and the self-referential paradoxes it gives rise to, is just what consciousness \emph{is}: ``In the end, we self-perceiving, self-inventing, locked-in mirages are little miracles of self-reference."} The sentence is contradictory because if the sentence is true, then because it asserts that it is false, so the sentence is false; if the sentence is false, then because its assertion must be negated, so the sentence is true. So the sentence can be neither true nor false. 

Perhaps you are thinking that the only thing more absurd than the absurd sentence is how it relates to culture. So let me explain.

\newpage
\section{Equilibrium, Mediation, and the Halting Problem}
\begin{quotation}
``G{\"o}del's theorems supply impressive arguments in favor of constructivism." 
\attrib{Jean Piaget}
\end{quotation}
In this section, I tie Piaget's equilibrum with uncomputability. I also show that Vygotskian mediation can be interpreted along the same line. 

Let us start with a brief primer on Turing machines. In 1936, Alan Turing shocked the world with his paper \emph{On Computable Numbers, with an Application to the Entscheidungsproblem}\cite{turing1936a}. In it, Turing invented a precise notion of computation, christened the Turing machine: a hypothetical computing machine that can model \emph{any} algorithmic computation.\footnote{Note that it captures only \emph{algorithmic} computation. Non-algorithmic computation, such as interactive computation, cannot be captured by a Turing machine. It is instructive to think of a Turing machine, and algorithmic computation, as a concretization of a priori rationality, whereas non-algorithmic computation can be thought of as a concretization of empiricism. For more, see \cite{Wegner}.} Intuitively, a Turing machine is an idealization of a person, equipped with a pen and unlimited pieces of paper, following a set of explicit rules, scribbling on the paper, until some result has been derived. The key idea is that the rules are explicit and finite, so that it can be written down and looked up whenever one wants to. A mathematical proof is also explicit and finite, so a computation by a Turing machine is equivalent to a mathematical proof. There are many ways of looking at a Turing machine. One is to consider it as an algorithm. Another is to consider it as a decider, deciding whether a given mathematical proposition is true or false. Yet another is to consider it as a Piagetian schema. I claim that all three perspectives are equivalent. To do so, I must make an assumption, which I wish to make loud and clear right here:

\emph{Assumption: the Church-Turing thesis is true. That is, anything that is physically computable is computable by some Turing machine. In other words, every person is at most a Universal Turing machine.}\footnote{Disclaimer: again, the thesis applies only for \emph{algorithmic}, a-priori-rationality-capturing computation; see Wegner 1997 for how the Persistent Turing Machine, which captures non-algorithmic computation, violates the thesis. It should be noted that the disclaimer does not weaken my argument in any way.}

The thesis seems drastic, but it is really nothing more than a commitment to a priori rationalism, and a commitment to the idea that a Turing machine can model any process of rational reasoning. In other words, it is an concretized definition of rationality, where we define a process as rational if and only if there exists a Turing machine that describes that process. This commitment to a priori rationalism gets us to say with perfect confidence that all people, insofar as they are free, are uncomputable and therefore equal. This is by way of Rice's theorem, an extension of the halting problem. What is the halting problem?

The halting problem asks, if there is a Turing machine that can decide whether all Turing machines will halt, or not. Another way of formulating the question is: is there any explicit, logically consistent procedure one can execute to decide whether all explicit, logically consistent procedures will ever end? As Turing showed, this problem is undecidable for much the same reason that the sentence ``this sentence is false" is neither true nor false. The reason the sentence is neither true nor false is because the sentence can refer to itself. Similarly, the halting problem is undecidable because a Turing machine can refer to itself. Consider this. A Turing machine is a sequential set of explicit instructions. The instructions explain what is to be done to some input. For example, ADDITION is a Turing machine which takes in two numbers as inputs, manipulates the inputs with some explicit instructions, and outputs their sum. The core idea is that there is no qualitative difference between the instruction and the input: they are both data. ADDITION, the sequential set of instructions, is just data, a bunch of 0's and 1's, just as the input is just data, a bunch of 0's and 1's. The equivalence of Turing machine and input permits using a Turing machine as an input to a Turing machine.

This idea is very much related to Piaget's invention and Vygotsky's reverse action/mediation. Piaget declares that the singular moment in the child's development is when s/he begins to invent. Vygotsky declares that the singular moment in the child's development is when s/he begins to perform reverse action. I claim that the two are saying the exact same thing, just in different words. Let's say that to invent is to act in such a way that could never have been predicted, that it is logically impossible for that action to have been predicted. Under this definition, I claim that reverse action enables invention. To define reverse action, an agent performs reverse action when the agent acts in such a way to modify itself. For example, a student writes something down, say a grocery list, and later performs actions according to what is written. This demonstrates the crucial equivalence of Turing machine and data: the student can output some data, i.e. the grocery list, that comes back and becomes a Turing machine, i.e. the shopping, for the student to perform. The idea that Turing machine and data are equivalent was the crucial realization that enabled computers to become so powerful. When computers were first invented, people designed separate machines for separate tasks. A calculator only did calculation; a text-editor only did text-editing. This was because they did not know the equivalence of Turing machine and data. A Turing machine, they thought, had to be some physical machine that looked at some data and output some data. Because of dichotomy of machine and data, because they did not realize the machine could be seen as just data, separate physical machines had to be designed for separate programs. Then, after realizing that Turing machines can be represented as data, the universal machine was invented. The universal machine could look at any Turing machine, represented as data, and do whatever that Turing machine would do. In this way, the universal machine could simulate any other Turing machine. The invention of the universal machine freed humanity from the burden of physically building -- with hands -- the horror! -- -- a separate machine for each separate task, and computers proliferated. This is evidenced by the smartphone, a universal machine, on your desk.

But the equivalence of Turing machine and data, as explained above, permits using a Turing machine as an input to a Turing machine, which in turn permits self-reference. And self-reference, like the sentence "this sentence is false", inevitably causes contradiction. To tie this back to Vygotsky and Piaget, I claim that the moment a child starts performing reverse action is when s/he realizes the equivalence of Turing machine and data. The moment a child starts performing reverse action is when he becomes capable of self-reference. Because of the contradiction inherent in this act, this is also the moment where others become incapable of predicting the child's every action -- the child becomes capable of invention. 

So the commitment to the Church-Turing thesis, or the commitment to a priori rationalism, gets us to say with perfect confidence that all people, insofar as they are free, are uncomputable and therefore equal. This is by way of Rice's theorem, an extension of the halting problem. But this reasoning gets stuck under the plain ugly empirical evidence that not everyone is uncomputable; in fact, they readily are. A racist who bristles at the sight of an purple person, or a sexist who gets mad at the sight of a woman CEO, are completely computable in the sense that the racist cannot help but bristle at the sight of a purple person, i.e. he is literally not free to not bristle. Ditto for the sexist. I define racism, or sexism, or the various other destructive -isms, as the axiomatic, unexamined, false belief that some syntactic property of a Turing machine (i.e. color of its skin) implies some semantic property of the Turing machine (i.e. whether the person is honest, capable of leadership, etc.). By Rice’s Theorem, no Turing machine, i.e. no rational process of reasoning, can decide a semantic property of an arbitrary Turing machine; therefore, racism or sexism is not just wrong but \emph{false}. 

Now let us intertwine Turing and Piaget. Piaget's grand schema is equilibrium: we assimilate what is in the world into schemas we have, and when we find our schemas are inadequate, we accommodate our schemas to fit the world. To make the terms precise, assimilation is ``the process by which the individual deals with an environmental event in terms of current structures", while accommodation is ``the individual's tendency to change in response to environmental demands" (Ginsburg \& Opper 1988). I find this formulation especially elegant, and there are numerous experimental and neuroscientific evidence backing it. In Piaget's words, a schema is ``a cohesive, repeatable action sequence possessing component actions that are tightly interconnected and governed by a core meaning" (Piaget 1952). Piaget wished to formalize psychology with precisely defined notions, and as such, his definition of a schema reads a lot like that of an algorithm: ``the set of rules a machine follows to achieve a particular goal", as defined by Merriam-Webster. If a Piagetian schema is equivalent to an algorithm, then it is natural to think of a schema as a Turing machine, because a Turing machine is just a concrete formulation of an algorithm. 

``Piaget reasoned the expected outcomes for each stage [of development] could be formalized in a kind of behavioral ``truth table" listing all of the possible transformations"\cite{burman}. Switching gears from Turing machines to Piagetian schemas, a true proposition is an action that can be performed, whereas a false proposition is an action that cannot be performed. In other words, a schema is a coherent logical structure that says whether some proposition is true or false, and a trut proposition translates into a performable action. As Burman explains, Piaget thought: one could write out all the possible propositions\footnote{A proposition can be written as a sequence of 0's and 1's. Therefore, there are $2^n$ propositions of length $n$. As explained in the previous section, this is an intractable number for even modestly big values of $n$, but it is still finite, so in principle, all the possible propositions can be written out.}, run all Turing machines (schemas) that a child of some stage is hypothesized to have on all of the propositions, and by looking at whether the Turing machine decides that the proposition is true or false, decide exactly what kinds of actions a child in some stage of development can and cannot do! 

But even ignoring the practical intractability, this is impossible even in principle. The reason: G{\"o}del's Incompleteness Theorem, as Piaget came to realize (Burman 2016).\footnote{G{\"o}del's Incompleteness theorem and the halting problem are two different formulations, one mathematical and one computational, of the exact same idea.} Some propositions, like the halting problem, or any proposition about a semantic property of all Turing machines, are necessarily undecidable. No Turing machine exists that decides a semantic property of all Turing machines. In other words, no schema exists that lets you draw watertight conclusions about a person, including yourself. A corollary is that, if you draw conclusions about a person, your schema has gone awry: it has run into a contradiction. What I am trying to say is something like racism or sexism, some axiomatic revulsion for certain kinds of people, is wrong, not just morally, but logically. Furthermore, \emph{if} you hold such axiomatic beliefs, you have fooled yourself. Holding such beliefs is simply impossible by the rules of logic; it can be proven that it cannot be proven. In this way, you have become the fool. You have become computable.\footnote{For a rigorous treatment of this argument, see Appendix A.}

But this fact spells trouble, because Piaget says that contradiction leads to disequilbrium. I have shown that fitting a person into a schema must lead to a contradiction; therefore, interacting with other humans must lead to disequilibrium. But the disequilibrium must be dispelled! How can this be done?

Piaget believed that humans strive towards equilibrium. This is true, but what I'm interested in is something he never seems to have focused on: \emph{dis}equilibrium. Clearly, the very notion that humans \emph{strive} towards equilibrium imply that they are at a state of disequilibrium, and it is my intuition that most people, except for the truly enlightened few, are almost always at a state of disequilibrium. Since equilibrium is an interaction between assimilation and accommodation, disequilibrium must then be an overabundance of one at the expense of the other: too much assimilation or too much accommodation. How do the disequilibria manifest? In the form of distinct reasoning patterns, which characterize distinct cultures.

\newpage
\section{The Body, Relational Reasoning, Exponentiation}
\begin{quotation}
``You have received your body, skin, and hair from your father and mother, therefore to keep them pristine is the start of filial piety."

\attrib{Confucius}
\end{quotation}
Here is an experiment I have done on babies. First, the baby is seated in front of a box. I put two identical wooden blocks on top of the box, and make the box play music. Then, I put two different wooden blocks on top of the box, and do not make the box play music. After the baby (typically 3-4 years old) has observed this, I give him/her two identical blocks and two different blocks, and ask, ``Can you make the box play music?" The point of the experiment is that the baby must reason about \emph{relations} between objects, not objects themselves, in order to find what it is that makes the box play music. What is interesting is that real baby babies, about 1$\frac{1}{2}$ years old, solve this puzzle easily, but relatively mature babies, about 3 years old, fail\cite{caren}.

English babies about 14 months old exhibit a noun spurt. Korean babies around the same age exhibit a verb spurt\cite{gopnik}. Since verbs encode relations between objects, an emphasis on verbs may place more salience in relational reasoning. So we had a hypothesis that Korean-speaking 3-year olds may be able to solve the aforementioned task more consistently than English-speaking 3-year olds. Over the past year, I did the task on about a hundred Korean babies, and a few English babies. Frequently, English babies would say stuff like ``the blue block plays music!", showing their bias for finding salience in single entities over relations. \footnote{It is my own empirically unverified hypothesis\footnote{a verification is on the way, in the form of using dolls and animals instead of wooden blocks} that Korean babies must be anthromorphizing the blocks, thinking of the blocks as if they were people. Because in person from a typically Eatern cultureic cultures, the concept of a person is contingent upon the person's relations with other people, the anthromorphization must lead to the babies thinking of relations between the blocks.}Illuminatingly, I found that Korean babies in the city perform at the same level as English-speaking babies, whereas Korean babies in rural areas perform much better than their formidable city-dwelling counterparts.

This finding gave the hypothesis a twist. Babies in both areas speak Korean all the same, so the argument that the verb-centric nature of Korean \emph{causes} an emphasis on relational reasoning is untenable. So I concluded that the significant difference must be the degree to which the area has typically typically Western characteristics, in which it is insisted that the world be computable: more so in the city, less so in the rural areas.

Insisting that the world be computable leads to overassimilation, which puts undue salience on single entities. There are more relations between things than there are things, so it is easy to find the watertight closure of computability among single entities than it is with relations. But insisting on this closure leads to assimilating even such things that cannot be assimilated. It is a mathematical fact that given $n$ things, there are always $2^n$ subsets of the things. Given a set of things, a relation exists for each subset of the things: therefore, there are $2^n$ relations among $n$ things. $2^n$ becomes very big very fast as $n$ becomes big. In computational complexity theory, it is termed \emph{intractable}. This is the reason why games like chess and Go are so hard for computers to play: the number of possible moves grows exponentially, prohibiting a brute-force exhaustive search.\footnote{Suppose you are playing chess, you have 10 possible moves each turn, and the game takes 80 turns. Then there are $10^{80}$ total possible moves, which is about the number of atoms in the universe.} For the same reason, no algorithm can exist for each and every relation. In other words, relations cannot all be assimilated. There are just too many of them! To make it more concrete, among just three people, say Alice, Bob, and Caren, there are $2^3 = 8$ relations: the empty relation, the relation of Alice to herself, the relation of Bob to himself, the relation of Caren to herself, the relation between Alice and Bob, the relation between Alice and Caren, the relation between Bob and Caren, and the relation between Alice, Bob, and Caren.\footnote{If there an infinite number of things, the number of relations among those things is a higher-order infinity. Exponentiation by a finite number leads to intractability; exponentiation by an infinite number leads to downright uncomputability.} So precise, algorithmic reasoning over relations is intractable, and for all intents and purposes, impossible. An emphasis on relational reasoning is therefore incompatible with the closure of computability. 

Then how is relational reasoning done? There are many potential candidates. One obvious candidate is the use of embodied metaphors. Lakoff and Johnson\cite{metaphors-we-live-by} claim that almost all human reasoning is done through the use of embodied metaphors instead of rational reasoning. This idea is consistent with empirical neuroscientific evidence that verbs activate the area in the brain that does motor control\cite{foroni}. It is also consistent with my observation that the body is just generally held in higher regard in Korea than in America.\footnote{One of the weirdest things about American culture, I have observed, is that disregard for the body, being out of touch with the body, is seen as something to be desired, something that lets you ``not give a damn", something that lets you be more \emph{masculine}. It is my hunch that any given cluster of Koreans, or any person from a typically Eatern cultureic people, are unconsciously aware of one another's bodily states at all times. To facilitate such coordination, Koreans constantly communicate to each other about the state of their body, such as ``it's so cold", ``it's so hot", ``it's spicy", etc. (Ever watch the Chinese Olympic opening ceremony and wonder how the thousands of performers are so perfectly synchronized?) This is such a culturally engrained activity that I frequently attempt such communication about my bodily state in America without second thought, only to be faced with at best nonchalance and at worst emasculation.} As the quotation opening the section shows, in typically Eatern societies, the body is highly respected. This is not a nonexistent idea in typically Western societies, either, but an emphasis on single-entity reasoning, tractable to the closure of computability, does not require respect for the body, while relational reasoning is completely reliant on the body. Therefore, emphasis on relational reasoning goes hand in hand with high respect for the body, which is consistent with the fact that Korea has a verb-centric language, and that the semantics of verbs are grounded in bodily feelings and motor control. 

Here's a little speculation. In \emph{An Enquiry Concerning Human Understanding}, Hume talks about the distinction between remembering a mathematical statement and remembering the feeling of a searing burn. The distinction, he claims, is that a mathematical statement is the same freshness whether he reads it from a piece of paper or remembers it in his head, while the experience of a searing burn is very different from remembering the experience of a searing burn. How can this be so? I claim that it is because a schema exists for the mathematical statement whereas a schema does not exist for the feeling of a searing burn. To read or to remember the mathematical statement is to execute its corresponding schema. Because the execution of the schema is the same, the feeling evoked is the same. However, because no schema exists for the feeling of a searing burn, it simply cannot be precisely executed. That no schema exists is the same as saying that no Turing machine exists, which means that it is uncomputable. So, bodily sensations are uncomputable. Extending this line of thought, consider an image of an apple versus the sensation of walking. What is more readily imaginable? An image of an apple, of course. So the image of an apple is computable, while the sensation of walking is uncomputable. What does it mean that motor control is uncomputable, when we plainly can control our motors? I mean that it cannot be simulated in a schema; it cannot simply imagined and have the same affect as actually acting it out. Neuroscientific evidence\cite{foroni} tells us that verbs activate the region of the brain that does motor control. So verbs encode motor control, and motor control is uncomputable. This should imply that verbs encode the uncomputable, which makes sense because person from a typically Eatern cultureic cultures tend to use verb-centric language while typically Western cultures tend to use noun-centric language (Gopnik \& Choi, 1995).
\newpage

\section{Dialectical Thinking}
\begin{quotation}
``The logical ways of dealing with contradiction may be optimal for scientific exploration and the search for facts because of their aggressive, linear, and argumentative style. On the other hand, dialectical reasoning may be preferrable for negotiating intelligently in complex social interactions. Therefore, ideal thought tendencies might be a combination of the both -- the synthesis, in effect, of Eastern and Western ways of thinking. "
\attrib{Kaiping Peng}
``The more I think about language, the more it amazes me that people ever understand each other at all."
\attrib{Kurt G{\"o}del}
\end{quotation}
If to understand a person is to know the output of that person's schemas, understanding, logically speaking, is impossible! It can be logically proven that there is no logical way to deduce the outputs of all schemas, or Turing machines, of a person. This is just a paraphrase of Rice's theorem. Then how is it that we ever understand each other at all? I realize the question sounds absurd to someone who has lived in more or less the same culture their entire life. \emph{Of course human interaction is possible, I do it every day!} But my argument is talking about human interaction in terms of logic and only logic. What I really wish to say is that human interaction based on logic is impossible, therefore it must not be based on logic, therefore it must be full of contradictions. 

Support for this line of thought comes from Peng, a cultural psychologist at UC Berkeley. He is famous for his study on how the typical Easterner and the typical Westerner think differently. Specifically, he talks about ``dialectical reasoning", a sort of reasoning that tolerates contradictions, as the default reasoning pattern for Easterners\cite{Peng}. For example, consider the two contradictory sentences, ``a butterfly in Beijing flapped its wings and it caused a storm in San Francisco", and ``a butterfly in Beijing flapped its wings and it had no effect on San Francisco's weather." American subjects, when presented with the two contradictory sentences, tended to polarize into agreeing with just one. Chinese subjects, on the other hand, tended to say that both sentences had some degree of truth in them. 

Peng simply observes that this is a phenomenon that happens. I am asking what the root of the phenomenon is. My answer is this: typically Eastern cultures tend to be disequillibrated in the too-much-accommodation direction, while typically Western cultures tend to be disequillibrated in the too-much-assimilation direction.\footnote{Here is a cute example as to how exactly typically Eastern cultures underassimilate. To explain the concept of generalizing assimilation, Piaget discusses the infant's tendency for thumb-sucking. I was struck by this discussion, because thumb-sucking, even by infants, is highly frowned upon in Korea. Koreans will go to great lengths to make sure their babies do not suck their thumbs. I once read a blog post by a mom bragging about how she got her child to stop sucking her thumb by wrapping her hands with stockings. The baby looked very sad in the picture. This is hardly proof that generalizing assimilation is performed to a lesser degree in typically Eastern cultures, but I stand behind the claim. Once I opened a box with a pen -- generalizing assimilation by extending the schema of a pen towards objects conventionally not used with a pen -- and my mother scolded me, saying that a box must be opened with a knife. It is my conviction that in Korea, the function of a tool is strictly confined to what the tool is supposed to do. Generalizing assimilation is thus avoided.}

To be part of a typically Eastern culture is to think about other people more, and to define oneself in terms of other people. There is neuroscientific evidence that, for a person from a typically Eastern culture, the part of the brain that lights up when thinking of the ego -- the frontal cortex -- is activated whether he is thinking of himself or his mother\cite{Kitayama}. No such phenomenon happens for a person from an typically Western culture. For the person from a typically Eatern culture, one's identity is dependent on relations between other people. In other words, the person from a typically Eatern culture undergoes constant accommodation until his desire is finally fulfilled when he meets his expected role in society. Complex, prescriptive social roles must exist in a person from a typically Eatern cultureic society. Without it, there is literally nothing to live for, nor to make sense of. The complex social prescriptions are arbitrary axioms, taken as given, because there is no point in arguing about them with logic.

Another little anecdote: when I was young, I annoyed my parents with an incessant stream of ``why"'s. This is probably the same for any petulant child, but the annoyance I felt was something different: a hushed, primeval shushing, like they did not even want to think about the ``why". This is because of the very respectable belief that some things must be left as axioms, that some ideas must be simply there and everyone must agree with them if we are to have a society that makes any ounce of sense. Respecting your elders is one of them. Receiving gifts with two hands, not one, is another. There is no ``why" to these rules because they are axioms. Without them, no complex social structure can exist. person from a typically Eatern cultureic culture is dependent on axioms like this because it is impossible to deduce all intents of another person. An illuminating example is this popular phrase in Korea, ``if each of us guard/follow what we will guard/follow, there is no misunderstanding and it is good." The phrase plainly states what I just described: axioms are axioms and we must follow what they say because otherwise there is no way we can understand each other. This brings to mind Vygotksy's theory of social meaning. Vygotsky gives an example where an infant learns the meaning of grasping because when he reaches out towards an object his mother fetches it for him. The meaning is constructed because of the mother's action. I think this is a good general framework to think about how these axioms come to exist. For example, it is rude to receive a gift with just one hand, and it is similarly rude to hand over a knife to someone while grabbing its handle. One is expected to receive a gift with two hands, and to hand over a knife while grabbing the blade side and having the handle point to the receiver. The rudeness is not inherent; it is constructed by someone in the culture's reaction.

On the other hand, in an person from a typically Western cultureic society, social roles tend to be less complex and more freewheeling. In an typically Western society, identity is about oneself and not about relations. An typically Western considers himself complete and people computable. There is no reason to obsess over relations. As such, the idea of just what a human being is is radically different in person from a typically Eatern cultureic versus typically Western cultures. In Korea, the word for human literally means ``between people" -- it is the betweenness, the relationships the person has, that defines a person. It is a common Korean mood that just because you are technically of the genus \emph{homo sapiens} does not qualify you to be a \emph{human being}. The narrative is that one is born an unenlightened beast, and life is a long and arudous process towards becoming a human being; in other words, the schemas one develops will never be enough and must go through constant accommodation.

In the previous section, we have argued that it is logically impossible to fit an entire person into a schema. And people from person from a typically Eatern cultureic cultures know this. They do not try to fit a person into a schema, and instead constantly accommodate the schema (there are exceptions -- the other, as opposed to the us, is very easily schematized). A side effect is that this line of thinking bleeds into places where it should not. While clearly there are a lot of things in the world that can be described by a coherent schema, such as a theory of physics, people from person from a typically Eatern cultureic cultures tend to be skeptical of this because they are so used to accommodation. In other words, because their default mode of computation is about the uncomputable, they mistake the computable as uncomputable. 

This is dangerous, for it leads to not believing in science, to suspicion towards rigorous reasoning. The Korean ruling elite disdained science and engineering for half a millenia, considering it childish. But what can be assimilated should be assimilated. If a schema can be developed, it should be developed. Once a schema of rigorous logical reasoning has been developed, one starts to see that all humans are equal. This is not possible if, in the person from a typically Eatern cultureic society disdaining the computable, there is an axiom that declares a whole swath of people as computable. The hallmark of person from a typically Eatern cultureic society is its amazing warmth towards insiders and equally amazing hostility towards outsiders. Without rigorous logical reasoning, this dichotomy can never be resolved.

People from typically Western cultures are stronger on this front: they are very wary of contradictions and know that a lot of things in the world can be described by a coherent schema. However, this line of thinking bleeds into places where it should not. This leads to the conception of a person as something that can be computed. Thus, to the person from a typically Western culture, a person is an object, a thing that can be simulated in a schema. 

This is dangerous, for it leads to dehumanization, to the lack of respect towards basic human decency. It is my lived experience that a person, simply by virtue of being a person, is afforded some baseline degree of respect in Korea. People are regularly outraged at something like a video of a company employee talking down to the janitor. The police do not carry guns and are as polite as can be. (Again -- as long as you are in the in-group.) This basic respect, I believe, is missing in typically Western cultures. And how could it not be, if one believes a person is computable? That something is computable means that it can be simulated however many times. If something is computable, it has no intangible value. It is, in the very precise sense of the word, an object.

The different conceptualizations on the game rock, paper, scissors illustrates this perfectly. To the person from a typically Eatern culture, rock, paper, scissors is just about the fairest game one could think of. In Korea, even adults play the game very often. The game is a recognition of each other's uncomputability, an affirmation that we are all free. The randomness of the game is what the person from a typically Eatern cultures seek. In America, I have been severely creeped out by my opponents who try to predict my next move in rock, paper, scissors: this is precisely the opposite of what the game's point in Korea! To the American who tries to predict his opponent's move in rock, paper, scissors, the game is about who can compute who better. It starts with the assumption that humans are computable, that a human can readily fit into a schema. For this reason, I believe to predict a move in rock, paper, scissors is nothing short of evil.
\newpage
\section{Conclusions}
For the person from a typically Western culture, the default is to assume the environment is computable, and to assimilate the environment; for the person from a typically Eatern culture, the default is to assume the environment is uncomputable, and to accommodate to the environment. For the person from a typically Western culture, the assumption of computability bleeds into objects that are uncomputable, such as humans; for the person from a typically Eatern culture, the assumption of uncomputability bleeds into objects that are computable, causing superstition. To be free, one must stop the bleeding. One who computes exactly what can be computed and does not compute exactly what cannot be computed is free. The conclusion is more stark if one consides the Kantian idea that to be free is to be moral. To compute exactly what one can compute, and to not compute exactly what one cannot compute, is to be uncomputable; to be uncomputable is to be free; to be free is to be moral. Therefore, the problem of morality is reduced to the ``synthesis, in effect, of Eastern and Western ways of thinking" (Peng). This synthesis allows the banishment of Piagetian disequilbrium, putting one on the fast lane to the ever-rising hymn of equilibrium, or equivalently, being uncomputable. Thus the is-ought problem is resolved. 

The lesson is that equilibration must go on. One must constantly strive for a state of equilibrium; one must not consider oneself capable of computing what is uncomputable, and one must not consider the oneself incapable of computing what is computable. One must not accommodate too much nor assimilate too much: a balance must be held. person from a typically Eatern cultures have a tendency to accomodate too much, to consider uncomputable what is computable; person from a typically Western cultures have a tendency to assimilate too much, to consider computable what is uncomputable. Both deficiencies make them not free. Both must learn something from the other. 

\newpage

\bibliographystyle{unsrt}
\bibliography{citations}
\end{document}